\documentstyle[12pt]{article}

\advance \topmargin by -\headheight
\advance \topmargin by -\headsep

\evensidemargin \oddsidemargin
\begin{document}

%

\def\rf#1{(\ref{eq:#1})}
\def\lab#1{\label{eq:#1}}
\def\nonu{\nonumber}
\def\br{\begin{eqnarray}}
\def\er{\end{eqnarray}}
\def\be{\begin{equation}}
\def\ee{\end{equation}}
\def\lb{\lbrack}
\def\rb{\rbrack}
\def\llangle{\left\langle}
\def\rrangle{\right\rangle}
\def\blangle{\Bigl\langle}
\def\brangle{\Bigr\rangle}
\def\llbrack{\left\lbrack}
\def\rrbrack{\right\rbrack}
\def\lcurl{\left\{}
\def\rcurl{\right\}}
\def\({\left(}
\def\){\right)}
\def\v{\vert}                     
\def\bv{\bigm\vert}               
\def\Bgv{\;\Bigg\vert}            
\def\bgv{\bigg\vert}              
\def\lskip{\vskip\baselineskip\vskip-\parskip\noindent}
\relax


\def\tr{\mathop{\rm tr}}                  
\def\Tr{\mathop{\rm Tr}}                  
\def\partder#1#2{{{\partial #1}\over{\partial #2}}}
\def\funcder#1#2{{{\delta #1}\over{\delta #2}}}


\def\a{\alpha}
\def\b{\beta}
\def\d{\delta}
\def\D{\Delta}
\def\eps{\epsilon}
\def\vareps{\varepsilon}
\def\g{\gamma}
\def\G{\Gamma}
\def\grad{\nabla}
\def\h{{1\over 2}}
\def\l{\lambda}
\def\L{\Lambda}
\def\m{\mu}
\def\n{\nu}
\def\o{\over}
\def\om{\omega}
\def\O{\Omega}
\def\p{\phi}
\def\P{\Phi}
\def\pa{\partial}
\def\pr{\prime}
\def\ra{\rightarrow}
\def\s{\sigma}
\def\S{\Sigma}
\def\t{\tau}
\def\th{\theta}
\def\Th{\Theta}
\def\z{\zeta}
\def\ti{\tilde}
\def\wti{\widetilde}
\def\suma#1{\sum_{#1}^{\infty}}   

\def\phanta{\phantom{blablablablabla}}
\def\phantb{\phantom{blablablablablablablablablab}}
\def\phantc{\phantom{blablablablablablablablablablablablablabla}}


\def\lie{{\cal G}}
\def\dlie{{\cal G}^{\ast}}
\def\elie{{\widetilde \lie}}
\def\edlie{{\elie}^{\ast}}
\def\hlie{{\cal H}}
\def\wlie{{\widetilde \lie}}
\def\f#1#2#3{f^{{#1}{#2}}_{#3}}                   


\font\numbers=cmss12
\font\upright=cmu10 scaled\magstep1
\def\stroke{\vrule height8pt width0.4pt depth-0.1pt}
\def\topfleck{\vrule height8pt width0.5pt depth-5.9pt}
\def\botfleck{\vrule height2pt width0.5pt depth0.1pt}
\def\Rmath{\vcenter{\hbox{\upright\rlap{I}\kern 1.7pt R}}}
\def\IR{\ifmmode\Rmath\else$\Rmath$\fi}



\def\Ouc{{\cal O}_{(U_0 ,c)}}           
\def\Gsu{G_{stat}(U_0 ,c)}                   
\def\Gs{G_{stat}}                            
\def\Asu{{\lie}_{stat} (U_0 ,c)}                   
\def\As{{\lie}_{stat}}                             
\def\Suc#1{\Sigma \Bigl( #1 ; (U_0 ,c) \Bigr)}        
\def\suc#1{{\hat \sigma} (#1 ; (U_0 ,c))}
\def\sh{\hat s}                              
\def\ssh#1{{\hat \sigma}^{#1}}
\def\Y#1{Y(#1)}
\def\y{{\hat y}}
\def\yp{y_{+}(g^{-1})}
\def\YT{Y_t (g^{-1})}
\def\yt{Y_t (g)}
\def\W1#1{W \lbrack #1 \rbrack}                
\def\Wuc#1{W \lbrack #1 ; (U_0 ,c)\rbrack}    
\def\Bil#1#2{\Bigl\langle {#1} \Bigg\vert {#2} \Bigr\rangle}  
\def\bil#1#2{\left\langle {#1} \bigg\vert {#2} \right\rangle} 
\def\me#1#2{\left\langle #1\right|\left. #2 \right\rangle}


\def\hd{{\widehat D}}
\def\dt{{\hat d}}
\def\Gpr{G^{\pr}}
\def\GT{\tilde \Gamma}
\def\GTy{\funcder {\GT} {y (t)}}
\def\GTz#1{\funcder {\GT} {y_{#1} (t)}}
\def\Ly#1{{\hat L}^{#1}_t (y)}
\def\Ry#1{R^{#1}_t (y)}
\def\LA{{\cal L}^A}


\def\Tu#1{{\widetilde \Theta}^{#1}}
\def\Td#1{{\widetilde \Theta}_{#1}}
\def\Z{\widetilde Z}
\def\T#1{{\hat {\cal T}}(#1)}
\def\dNz{\delta^{(N)} (z_1 - z_2 )}
\def\dNth{\delta^{(N)} (\th_1 - \th_2 )}
\def\dN#1#2{\delta^{(N)} ({#1} - {#2})}
\def\DN{{\llbrack D {\widetilde \Theta} \rrbrack}^2_N}
\def\Du#1{{\widetilde D}^{#1}}
\def\Dd#1{{\widetilde D}_{#1}}


\def\Tor{{\wti {\rm SDiff}}\, (T^2 )}
\def\Lh#1{{\hat {\cal L}}({#1})}
\def\M{{\cal M}}
\def\dM{{\cal M}^{\ast}}
\def\Mc{{\cal M}(R^1 \times S^1 )}
\def\dMc{{\cal M}^{\ast}(R^1 \times S^1 )}
\def\st1{\stackrel{\ast}{,}}


\def\Winf{{\bf W_\infty}}
\def\Win1{{\bf W_{1+\infty}}}
\def\Winft#1{{\bf W_\infty^{\geq {#1}}}}    
\def\DO{DOP (S^1 )}                           
\def\DA{{\cal DOP} (S^1 )}                    
\def\eDA{{\widetilde {\cal DOP}} (S^1 )}     
\def\dDA{{\cal DOP}^{\ast} (S^1 )}                  
\def\edDA{{\widetilde {\cal DOP}}^{\ast} (S^1 )}   
\def\DOP#1{{DOP (S^1 )}_{\geq{#1}}}                 
\def\DAP#1{{{\cal DOP} (S^1 )}_{\geq{#1}}}          
\def\eDAP#1{{{\widetilde {\cal DOP}} (S^1 )}_{\geq{#1}}}
\def\dDAP#1{{{\cal DOP}^{\ast} (S^1 )}_{\geq{#1}}}     
\def\edDAP#1{{{\widetilde {\cal DOP}}^{\ast} (S^1 )}_{\geq{#1}}}
\def\sto{\stackrel{\circ}{,}}              
\def\sta{\, ,\,}
\def\xx{(\xi , x)}
\def\yy{(\zeta , y)}
\def\xxt{(\xi , x ; t )}
\def\intres{\int dx\, {\rm Res}_\xi \; }
\def\intrest{\int dt\, dx\, {\rm Res}_\xi \;}
\def\Res{{\rm Res}_\xi}
\def\pexx{e^{\pa_x \pa_\xi}}
\def\mexx{e^{-\pa_x \pa_\xi}}
\def\SLinf{SL (\infty ; \IR )}             
\def\slinf{sl (\infty ; \IR )}               
\def\sumlm{\sum_{l=1}^{\infty} \sum_{\v m\v \leq l}}
\def\WDO#1{W_{DOP (S^1 )} \lb #1\rb}               


\def\PsDA{\Psi{\cal DO} (S^1 )}
\def\dPsDA{\Psi{\cal DO}^{\ast} (S^1 )}
\def\PsDO{\Psi {\rm DO} (S^1 )}
\def\Volt{\Bigl( \Psi{\cal DO} \Bigr)_{-}}
\def\dVolt{\Bigl( \Psi{\cal DO} \Bigr)_{-}}
\def\VOLT{\Bigl( \Psi {\rm DO} \Bigr)_{-}}
\def\Rm#1#2{r(\vec{#1},\vec{#2})}          
\def\OR#1{{\cal O}(R_{#1})}           
\def\ORti{{\cal O}({\widetilde R})}           
\def\AdR#1{Ad_{R_{#1}}}              
\def\dAdR#1{Ad_{R_{#1}^{\ast}}}      
\def\adR#1{ad_{R_{#1}^{\ast}}}       
\def\AdB#1{Ad\Bigl( g^{-1}_{#1}(L) \Bigr)}  
\def\dAdB#1{Ad^{\ast}\Bigl( g_{#1}(L) \Bigr)}  
\def\KP{${\bf \, KP\,}$}                 
\def\KPl{${\bf \,KP_{\ell}\,}$}         
\def\KPo{${\bf \,KP_{\ell = 0}\,}$}         
\def\mKPa{${\bf \,KP_{\ell = 1}\,}$}    
\def\mKPb{${\bf \,KP_{\ell = 2}\,}$}    

\def\A{\cal A}

\newcommand{\nit}{\noindent}
\newcommand{\ct}[1]{\cite{#1}}
\newcommand{\bi}[1]{\bibitem{#1}}
%
%
\def\PRL#1#2#3{{\sl Phys. Rev. Lett.} {\bf#1} (#2) #3}
\def\NPB#1#2#3{{\sl Nucl. Phys.} {\bf B#1} (#2) #3}
\def\NPBFS#1#2#3#4{{\sl Nucl. Phys.} {\bf B#2} [FS#1] (#3) #4}
\def\CMP#1#2#3{{\sl Comm. Math. Phys.} {\bf #1} (#2) #3}
\def\PRD#1#2#3{{\sl Phys. Rev.} {\bf D#1} (#2) #3}
\def\PLA#1#2#3{{\sl Phys. Lett.} {\bf #1A} (#2) #3}
\def\PLB#1#2#3{{\sl Phys. Lett.} {\bf #1B} (#2) #3}
\def\JMP#1#2#3{{\sl J. Math. Phys.} {\bf #1} (#2) #3}
\def\PTP#1#2#3{{\sl Prog. Theor. Phys.} {\bf #1} (#2) #3}
\def\SPTP#1#2#3{{\sl Suppl. Prog. Theor. Phys.} {\bf #1} (#2) #3}
\def\AoP#1#2#3{{\sl Ann. of Phys.} {\bf #1} (#2) #3}
\def\PNAS#1#2#3{{\sl Proc. Natl. Acad. Sci. USA} {\bf #1} (#2) #3}
\def\RMP#1#2#3{{\sl Rev. Mod. Phys.} {\bf #1} (#2) #3}
\def\PR#1#2#3{{\sl Phys. Reports} {\bf #1} (#2) #3}
\def\AoM#1#2#3{{\sl Ann. of Math.} {\bf #1} (#2) #3}
\def\UMN#1#2#3{{\sl Usp. Mat. Nauk} {\bf #1} (#2) #3}
\def\RMS#1#2#3{{\sl Russian Math Surveys} {\bf #1} (#2) #3}
\def\FAP#1#2#3{{\sl Funkt. Anal. Prilozheniya} {\bf #1} (#2) #3}
\def\FAaIA#1#2#3{{\sl Functional Analysis and Its Application} {\bf #1}
(#2) #3}
\def\TAMS#1#2#3{{\sl Trans. Am. Math. Soc.} {\bf #1} (#2) #3}
\def\Invm#1#2#3{{\sl Invent. math.} {\bf #1} (#2) #3}
\def\LMP#1#2#3{{\sl Letters in Math. Phys.} {\bf #1} (#2) #3}
\def\IJMPA#1#2#3{{\sl Int. J. Mod. Phys.} {\bf A#1} (#2) #3}
\def\AdM#1#2#3{{\sl Advances in Math.} {\bf #1} (#2) #3}
\def\APP#1#2#3{{\sl Acta Phys. Polon.} {\bf #1} (#2) #3}
\def\TMP#1#2#3{{\sl Theor. Mat. Phys.} {\bf #1} (#2) #3}
\def\JPA#1#2#3{{\sl J. Physics} {\bf A#1} (#2) #3}
\def\JSM#1#2#3{{\sl J. Soviet Math.} {\bf #1} (#2) #3}
\def\MPLA#1#2#3{{\sl Mod. Phys. Lett.} {\bf A#1} (#2) #3}
\def\JETP#1#2#3{{\sl Sov. Phys. JETP} {\bf #1} (#2) #3}
\def\PJAS#1#2#3{{\sl Proc. Jpn. Acad. Sci.} {\bf #1} (#2) #3}
\def\JPSJ#1#2#3{{\sl J. Phys. Soc. Jpn.} {\bf #1} (#2) #3}
\def\JETPL#1#2#3{{\sl  Sov. Phys. JETP Lett.} {\bf #1} (#2) #3}

\begin{titlepage}
\vspace*{-1cm}
\noindent
\hfill{CERN-TH.6627/92} \\
\phantom{bla}
\hfill{UICHEP-TH/92-13} \\
\phantom{bla}
\hfill{hep-th/9209006}
\\
\vskip .3in
\begin{center}
{\large\bf R-Matrix Formulation of KP Hierarchies and Their
Gauge Equivalence}
\end{center}
\vskip .3in
\begin{center}
{ H. Aratyn\footnotemark
\footnotetext{Work supported in part by U.S. Department of Energy,
contract DE-FG02-84ER40173}}
\par \vskip .1in \noindent
Department of Physics , University of Illinois at Chicago\\
Box 4348, Chicago, Illinois 60680, U.S.A. , {\em e-mail} :
u23325@uicvm \\
\par \vskip .3in
{ E. Nissimov$^{\,2}$  and S. Pacheva \footnotemark
\footnotetext{On leave from : Institute of Nuclear Research and Nuclear
Energy, Boul. Trakia 72, BG-1784 $\;$Sofia, Bulgaria. }}
\par \vskip .1in \noindent
Department of Physics, Ben-Gurion University of the Negev \\
Box 653, IL-84105 $\;$Beer Sheva, Israel \\
{\em e-mail} : emil@bguvms , svetlana@bguvms \\
and \\
Theory Division, CERN, CH-1211 $\;$Geneva 23, Switzerland \\
{\em e-mail} : nissimov@cernvm , svetlana@vxcern \\
\par \vskip .3in
I. Vaysburd
\par \vskip .1in \noindent
Racah Institute of Physics, Hebrew University\\
IL-91904 $\;$Jerusalem, Israel, {\em e-mail} : igor@hujivms  \\
\par \vskip .3in
\end{center}

\begin{abstract}

The Adler-Kostant-Symes $R$-bracket scheme is applied to the algebra of
pseudo-differential operators to relate the three integrable hierarchies:
KP and its two modifications, known as nonstandard integrable models.
All three hierarchies are shown to be equivalent and
connection is established in the form of a symplectic gauge transformation.
This construction results in a new representation of the W-infinity algebras
in terms of 4 bosonic fields.
\end{abstract}

\vfill{ \noindent
CERN-TH.6627/92
\newline
\noindent
{August 1992}}
\end{titlepage}
\noindent
{\large {\bf 1. Introduction}}
\lskip
One of the important and still unsolved problems of the two-dimensional
physics is to describe consistently
systems with infinitely many functional (field-) degrees of freedom.
Among such systems the largest attention was attained by the
Kadomtsev-Petviashvili ($KP$) completely integrable hierarchy which
proved to be relevant for a variety of the physical problems.
A recent and intriguing development in this field
is the appearance of the integrable hierarchies, including $KP$,
in the matrix models known to describe at multicritical points
$c \leq 1\,$  matter systems coupled to $D=2$ quantum gravity \ct{2d}.
In particular, the partition function of both discrete matrix models
\ct{MaMiMo,K} and of continuum $c \leq 1$ string field theory
\ct{DVV,FKN}
is expressed in terms of a constrained $\t$-function of the $KP$
hierarchy. The coupling constants in the matrix model partition function,
corresponding to various possible deviations from the critical points,
are the evolution parameters in the $KP$ hierarchy.

The most essential feature of the integrable hierarchies with infinite
number of degrees of freedom , which prompts their connection to $2D$
conformal field theories, $c\leq 1$ strings and their matrix model
counterparts, is their Hamiltonian structure \ct{FT87,dickey}. Thus, the
centerless Virasoro algebra provides the second Hamiltonian structure of
Korteweg-de-Vries (KdV) hierarchy. This is the algebra of constraints on
the partition function of
the one-matrix model \ct{2d}. In the same way $W_N$ algebras
\ct{zamo}, or more precisely, their semiclassical analogs -
the Gelfand-Dickey algebras \ct{GD}, give the second Hamiltonian
structures  of the generalized KdV hierarchies. $W_N$ are the algebras of
constraints on the partition function in multi-matrix and Kontsevich
matrix models \ct{MaMiMo,K}.
Finally, $\Win1\,$ algebra \ct{pope}, which is isomorphic
to the Lie algebra of differential operators on the circle
$\DA \,$ \ct{radul}, yields the first Hamiltonian structure of $KP$ hierarchy
\ct{watanabe,BAK85,wu91}.

An important open problem of the matrix model formulation of $2D$
quantum gravity is how to describe the interpolation between two
different vacua (one, characterized by $(p,q)$ conformal matter
and another one - by $(p^{\pr}, q^{\pr})$ , $p^{\pr} \neq p \, ,\,
q^{\pr} \neq q$). It is known \ct{DSSh} that this cannot be solved
in terms of the ordinary $KP$ evolution parameters.
A hope to solve this problem is to analyze all integrable
hierarchies which are equivalent to $KP$ . This is also an interesting
mathematical problem by its own value.

In this letter we would like to contribute to this
program. The powerful Adler-Kostant-Symes (AKS) scheme
for Lie-algebraic construction of
integrable models (substantially improved and extended by Reyman
and Semenov-Tian-Shansky) \ct{AKS} is applied to the algebra of
pseudo-differential operators on the circle $\PsDA\,$ \ct{treves}.
This scheme permits the treatment of the three different integrable
$KP$-like hierarchies -
ordinary $KP$ and its two modifications , known as nonstandard
integrable models \ct{BAK85}, on equal footing. They correspond to the
three possible splittings of the algebra $\PsDA\,$ into a linear sum of two
subalgebras. The main result is that all three hierarchies are proved to
be ``gauge" equivalent via a generalized Miura transformation.
The (field-dependent) ``gauge" transformations, which are explicitly
constructed, belong to well-defined subgroups of the formal group
of pseudo-differential operators -
the abelian group of operators of multiplication by a
function and the group of diffeomorphisms on the circle (Virasoro
group), respectively.
For the first of the modified $KP$ hierarchies, this ``gauge" equivalence was
previously established \ct{BAK90} in the ``semiclassical"
limit of $KP$ (known in nonlinear hydrodynamics as Benney integrable
hierarchy \ct{LM79}).

As an important byproduct, the above ``gauge" transformations together
with the AKS $R$-bracket scheme provide new explicit realizations of
$\Win1\,$ algebras in terms of an unconventional set of four Bose fields.
\lskip
{\large {\bf 2. The Adler-Kostant-Symes Scheme and Applications to
\KP $\,$ Hierarchies}}
\lskip
{\bf 2.1 General Scheme and ${\bf R}$ Operators}

It is well-known that the Lie algebra methods
allow for an unifying treatment of integrable systems \ct{FT87}.
One of the main purpose of this paper is to describe a relation between the
Lax formulation of various $KP$-type systems defined below and the $R$-operator
approach \ct{STS83,GKR88} to the integrable systems based on the AKS
scheme \ct{AKS}. We first recall the notion of integrability.
\lskip
{\sl Complete Integrability:}
Consider a Hamiltonian system with $n$ degrees of freedom possessing
standard Hamiltonian structure with Hamiltonian $\, H(p,q)$ and Poisson
bracket $\{ \cdot,\cdot \}$.
A Hamiltonian system is called completely (or Liouville) integrable
if it has $n$ conserved quantities (integrals of motion) $I_k (p,q)$
, $k=1, \ldots n$, which are in involution: $ \{ I_i , I_j \} = 0 $.
For such a system we can find the action-angle variables and write the general
solution to the equations of motion.
\lskip
{\sl Lax formulation:}
For infinite-dimensional integrable Hamiltonian systems
there exists the convenient Lax (or ``zero-curvature") formulation
\ct{FT87}.
In the Lax formulation the dynamical equations of motion can be written in
terms a Lax pair $L$, $P$, with values in some Lie algebra $\lie$, as
the Lax-type equation
\be
{ d L \o dt} = \lb L\, , \, P \rb
\lab{laxeq}
\ee
The Lax formulation leads straightforwardly to construction of the integrals
of motion. Namely, for any Ad-invariant function $I$ on $\lie$, $I \( L\) $
is a constant of motion.
In fact, it can be shown that any completely integrable Hamiltonian system
admits a Lax representation (at least locally) \ct{BV90}.
\lskip
{\sl The $KP$ hierarchy:} An important example of integrable system admitting
the Lax formulation is given by $KP$ hierarchy consisting of the following
family of Lax equations
\be
{ \pa L \o \pa t_r } = \lb L \, , \, L^r_{+} \rb \, , \qquad \quad r =1,2,3,
\ldots,
\lab{kpsystem}
\ee
where $L$ is a pseudo-differential operator
\be
L = D + \sum_{i = 1}^{\infty} u_i D^{-i}
\lab{kpl}
\ee
The subscript $(+)$ means taking the purely differential part of $\,
L^r \,$ and $t = \{ t_r \} $ are the evolution parameters
(infinitely many time coordinates).
The flows \rf{kpsystem} are bi-Hamiltonian \ct{dickey},
i.e. there exist two
Poisson bracket structures $\{\cdot , \cdot \}_{1,2}$ , such that
we can rewrite \rf{kpsystem} as :
\be
\partder {L} {t_r} = \{ H_r \,, \, L \}_2= \{ H_{r+1} \,, \, L \}_1
\lab{uflow}
\ee
Here the Hamiltonians for the $KP$ hierarchy are $H_r = { 1 \o r} \int
{\rm Res} \, L^r\,$ ( Res $\,$denotes the coefficient in front of
$\, D^{-1}$ ). The second equality in \rf{uflow} is a particular case
of the so called Lenard relations
$\{ H_{r+1}\, ,\, L  \}_m = \{ H_r \, ,\, L \}_{m+1}$
for a hierarchy of Poisson bracket structures $\, m=1,2, \ldots$.
There exists a fundamental theorem (see e.g. \ct{fordy}) connecting notion
of integrability with the property of possessing a bi-Hamiltonian structure,
which establishes the $KP$ system as integrable.
\lskip
{\sl The AKS Scheme:}
A very wide class of integrable models can be constructed through the
application of the AKS method having roots in the
coadjoint orbit formulation.

Let $G$ denote a Lie group and $\lie$ be its Lie algebra.
$G$ acts on $\lie$ by the adjoint action:
$ Ad(g) \, X = g X g^{-1}$ with $ g \in G$ and $X \in \lie$.
Let $\dlie$ be the dual space of $\lie$
relative to a non-degenerate bilinear form $\langle \cdot \v
\cdot \rangle$ on $\dlie \times \lie$.
The corresponding coadjoint action of $G$ on $\dlie$ is obtained from
duality of $\langle \cdot \v \cdot \rangle$ : $ \langle Ad^{\ast}(g) U
\v X \rangle = \langle U \v Ad (g^{-1}) X \rangle $ .
We will denote the infinitesimal versions of adjoint and coadjoint
transformations by $ad (Y)$ and $ad^{\ast} (Y)$ (for $ g = \exp Y$ ).

There exists a natural Poisson structure on the space
$C^{\infty} \(\dlie, \IR \)$ of smooth, real-valued functions on $\dlie$
called Lie-Poisson (LP) bracket.
The LP bracket for $F, H \in C^{\infty} \(\dlie, \IR \)$
is given by :
\be
\{ F \, , \, H \} (U) =
- \biggl\langle\, U \, \bigg\v \, \biggl\lb \nabla F (U) \, , \,
\nabla H (U) \biggr\rb \, \biggr\rangle
\label{eq:LBKKbra}
\ee
where the gradient $\nabla F\, : \, \dlie \ra \lie$ is defined by the
standard formula ${d \o dt} F \left( U + t V \right)\vert_{t=0} =
\llangle V \, \vert \, \nabla F (U) \rrangle$
and where $\lb \cdot \, , \, \cdot \rb $ is the standard Lie bracket
on $\lie$.
It follows clearly that $\{ \cdot \, ,\, \cdot \}$ is antisymmetric and it is
also easy to verify the Jacobi identity.
On each orbit in $\dlie$ the LP bracket gives rise to a
non-degenerate symplectic structure.
Moreover, for any Hamiltonian $H$ on such orbit we have a Hamiltonian
equation $ d U / dt = ad^{\ast} (\nabla H (U)) U$.

We now introduce $R$-operator (generalized
$R$-matrix) as a linear map from a Lie algebra $\lie$
to itself such that the bracket :
\be
\lb X , Y \rb_{R} \equiv \h \lb R X , Y \rb + \h \lb X, R Y \rb
\lab{R-lie}
\ee
defines a second Lie structure on $\lie\,$ \ct{STS83}.
The modified Yang-Baxter equation (YBE) for the $R$-matrix must hold in
order to ensure the Jacobi relation.

We can furthermore introduce a new LP bracket $\{ \cdot, \cdot\}_{R}$
called $R$-bracket substituting the usual Lie bracket
$ \lb \cdot , \cdot\rb $ by the $R$-Lie
bracket $ \lb \cdot , \cdot\rb_{R} $ \rf{R-lie} in \rf{LBKKbra}:
\be
\{ F \, , \, H \}_R (U) =
- \biggl\langle\, U \, \bigg\v \, \biggl\lb \nabla F (U) \, , \,
\nabla H (U) \biggr\rb_R \, \biggr\rangle
\lab{Rbra}
\ee
A function $H$ on $\dlie$ is called $Ad^{\ast}$-invariant
(Casimir) if $\; H\lb Ad^{\ast}(g) U\rb = H\lb U\rb\;$ or,
infinitesimally, $ ad^{\ast} (\nabla H (U)) (U) =0$ for each $U \in \dlie$.
Then one can show \ct{STS83} that : \\
\nit
(1) the $ad^{\ast}$-invariant functions are in involution
with respect to both brackets \rf{LBKKbra} and \rf{Rbra} ;\\
\nit
(2) the Hamiltonian equation on $\dlie$ takes the following
(generalized Lax) form :
\be
d U /dt = \h ad^{\ast} \Bigl(R \(\nabla H (U)\)\Bigr) U
\lab{rhameqs}
\ee
corresponding to the equations of motion $d F / dt = \{ H , F \}_{R}$ for
$F \in C^{\infty} \(\dlie, \IR \)$.

Hence the above R-matrix technique leads to a direct construction of
integrable systems based on Casimir functions on $\dlie$.
The basic realization of this technique arises when the Lie algebra $\lie$
decomposes as a vector space into two subalgebras $\lie_{+}$ and
$\lie_{-}$ i.e. $\lie = \lie_{+} \oplus \lie_{-}$. Let $P_{\pm}$ be the
corresponding projections on $\lie_{\pm}$. Then $R= P_{+} - P_{-}$
satisfies the modified YBE and provides a specific realization for the above
scheme.
\lskip
{\bf 2.2 AKS Construction of Three \KP Hierarchies}

Here we will illustrate the AKS construction on $\lie = \PsDA\,$ - the
Lie algebra of pseudo-differential operators on circle. Recall
that arbitrary
pseudo-differential operator $X (x, D_x ) = \sum_{k \geq -\infty}
X_k (x) D_x^k \,$ is conveniently represented by its symbol
\ct{treves} - a Laurent series in the variable $\xi$ :
\be
X \xx = \sum_{k \geq - \infty} X_k (x) \xi^k\lab{laurent}
\ee
and the operator multiplication corresponds to the following symbol
multiplication :
\be
X \xx \circ Y \xx = \sum_{N \geq 0} h^N {1 \o N !}
{\pa^N X \o \pa \xi^N} {\pa^N Y \o \pa x^N}
\lab{product}
\ee
which determines a Lie algebra structure given by a commutator
$\lb X , Y \rb \equiv {1\o h} \( X \circ Y - Y \circ X \)$.
Explicitly we have :
\be
\Bigl\lb X \xx \, , \, Y \xx \Bigr\rb= \sum_{N \geq 1} (h )^{N-1} {1 \o N !}
\left( {\pa^N X \o \pa \xi^N} {\pa^N Y \o \pa x^N} -
{\pa^N X \o \pa x^N} {\pa^N Y \o \pa \xi^N} \right)\lab{symbol}
\ee
The constant $h$ appearing in \rf{product} and \rf{symbol}
has the meaning of a deformation parameter and henceforth will be taken
$h=1$. The limit $ h \to 0$ defines the semiclassical limit
of $\PsDA\,$ where the Lie bracket \rf{symbol} reduces to a
two-dimensional Poisson bracket: $\lb X \xx , Y \xx \rb=
\(\pa X /\pa \xi\) \(\pa Y /\pa x\)-\(\pa X /\pa x \) \( \pa Y /\pa \xi\)$.

Using the Adler trace one next defines an invariant, non-degenerate
bilinear form :
\be
\blangle L \v X \brangle \equiv {\Tr}_A \( L X \) =
\intres L\xx \circ X\xx\lab{adler}
\ee
which allows an identification of the dual space $\dlie$ with $\lie$ and
the coadjoint action with the adjoint action.

There exist three natural decompositions of $\lie$ into a linear sum of two
subalgebras:
\be
\lie^{\ell}_{+} = \lcurl X_{+} \equiv X_{\geq \ell}
= \sum_{i=\ell}^{\infty} X_i (x)\xi^i\rcurl
\qquad;\qquad
\lie^{\ell}_{-} = \lcurl X_{-} \equiv X_{< \ell}
= \sum_{i=-\ell+1 }^{\infty} X_{-i}(x)\xi^{-i}
\rcurl\lab{subalg}
\ee
labeled by the index $\ell$ taking three values $\ell = 0,1,2$.
For each $\ell$ we clearly have $\lie = \lie^{\ell}_{+} \oplus
\lie^{\ell}_{-} $.
Correspondingly the dual spaces to subalgebras $\lie^{\ell}_{\pm}$ are given
by :
\be
{\lie^{\ell}_{+}}^{\ast} = \lcurl L_{-} \equiv L_{< -\ell}
= \sum_{i=\ell+1}^{\infty}
\xi^{-i} \circ u_{-i}(x)\rcurl \;\;\; ;\;\;\;
{\lie^{\ell}_{-}}^{\ast} = \lcurl L_{+} \equiv L_{\geq -\ell} =
\sum_{i=-\ell}^{\infty} \xi^{i} \circ u_{i}(x) \rcurl
\lab{dsubalg}
\ee
Note that in \rf{dsubalg} the differential operators are put to the
left.

Denoting by $R_{\ell} = P_{+} - P_{-}$ for each of three cases one finds
that $R$-bracket is given by :
\be
\lb X , Y \rb_{R_{\ell}} = \lb X_{\geq \ell},
Y_{\geq \ell}\rb - \lb X_{< \ell} , Y_{< \ell} \rb\lab{R-bra}
\ee
Furthermore, from the general relation for the $R$-coadjoint action of
$\lie\,$ on its dual space $\,ad^{\ast}_R (X) L =
\h ad^{\ast} (RX) L + \h R^{\ast} ad^{\ast} (X) L $ we find
the that the infinitesimal shift along an $R$-coadjoint orbit
$\OR{\ell}\,$ has the form :
\be
\d_{R_\ell} L \equiv
ad^{\ast}_{R_\ell} (X) L = \Bigl( ad^{\ast} \(X_{+} \) L_{-} \Bigr)_{-}
- \Bigl(ad^{\ast} \(X_{-}\) L_{+} \Bigr)_{+} \equiv
\left\lb X_{\geq \ell} , L_{< -\ell} \right\rb_{< -\ell}
- \left\lb X_{< \ell} , L_{\geq -\ell} \right\rb_{\geq -\ell}
\lab{R-coadj}
\ee
Henceforth, the subscripts $\,\pm \,$ will
denote projections on ${\lie_{\ell}}_{\pm}$ and
${\dlie_{\ell}}_{\mp}$ , as in \rf{subalg}, \rf{dsubalg} .
Also, we shall skip the sign $\circ$ in symbol products for
brevity.

We will now discuss in greater detail the Hamiltonian structure of the
integrable systems given by the three decompositions labeled by $\ell=0,1,2$
as defined by the AKS scheme with Hamiltonian equations of motion
\rf{rhameqs}. We will call the resulting hierarchies the $r$-$KP$ or
\KPl hierarchies.

\KPo : Here we take $R$-coadjoint orbit of the
form $\, \OR{0} = \Bigl\{ L =\xi + L_{-} \equiv \xi + \suma{k=1}
\xi^{-k} \, u_k (x) \Bigr\} \,$ .
Choosing as a Casimir function $H_{m+1} = { 1 \o {m+1}} \intres
L^{m+1}\,$ we get from \rf{rhameqs}:
\be
\partder{L}{t_m} = \h ad^{\ast} \Bigl( (\nabla H_{m+1})_{+} -
(\nabla H_{m+1})_{-} \Bigr)\, L
= ad^{\ast} \Bigl( (L^m )_{+} \Bigr) \, L \lab{rkp0}
\ee
with $ (L^m )_{+} = \sum_{j \geq 0} \bigl(
\d H_{m+1} / \d u_{j+1} (x) \bigr) \, \xi^j$ .
We recognize in \rf{rkp0} the standard $KP$ flow equation \rf{kpsystem}.
One finds the corresponding Hamiltonian structure to be induced by the
LP structure: $ \{ u_i (x) \, , \, u_j (y) \}_{R_0} = \O^{(\ell=0)}_{i-1,j-1}
(u(x))\,\d (x-y)$ where the form on the right hand side is given by
\ct{watanabe,BAK85}:
\be
\O^{(\ell)}_{i,j}\(u(x)\) = - \sum_{k=0}^{i+\ell}
{i+\ell\choose k} u_{i+j+\ell - k+1} (x) D^k_x +
\sum_{k=0}^{j+\ell} (-1)^k{j+\ell\choose k} D^k_x u_{i+j+\ell -k+1} (x)
\phantom{aa} \lab{watform}
\ee
for $\ell=0$. This LP bracket algebra is isomorphic to the centerless
$\Win1\,$ algebra \ct{wu91}. In conclusion we have found that
\KPo is the standard $KP$ hierarchy.

\mKPa : Here we first consider elements of
${{\cal G}_{-}^1}^{\ast} $ of the type $L_{+} = \xi + u_0 + \xi^{-1} \,
u_1\,$, which
preserve their form under $\d_{R_1} L_{+} = ad^{\ast}_{R_1} (X) L_{+}$
, i.e. they span an $R_1$-orbit of finite functional dimension $\, 2$ .
Calculation of the Poisson bracket according to \rf{Rbra} :
\be
\{ \me{L_{+}}{X} \, ,\,  \me{L_{+}}{Y} \}_{R_1} =
- \me{L_{+}}{\lb X\, , \, Y \rb_{R_1} }
\lab{rpb}
\ee
yields the $R$-brackets: $\{ u_0 (x) \, , \, u_1 (y) \}_{R_1}
= - \d^{\pr} (x-y)$ and zero otherwise.
We then define a complete Lax operator defined as
$ L^{(1)} = L_{+} + L_{-} = \xi + u_0 + \xi^{-1} \, u_1 +
\sum_{i \geq 2} \xi^{-i} \, v_{i-2}\,$.
Application of \rf{rpb} gives a Hamiltonian structure being a direct sum of
the matrix $P^{(1)}$ associated to the modes $\{ u_0, u_1 \}$ and the
Hamiltonian structure $\O^{(1)}$ associated to $\{ v_i \v i \geq 0\}$
\ct{BAK85} :
\be
\left(\begin{array}{cc}
P^{(1)}& 0\\
0& \O^{(1)}\end{array}
\right)
\;\; {\rm with} \;\;
P^{(1)}= \left(\begin{array}{cc}
0 & -\pa \\
- \pa & \; 0 \end{array}
\right)
\lab{ham1}
\ee
Note that $\O^{(1)}\,$ \rf{watform} corresponds to the centerless
$\Winf\,$ algebra.

\mKPb : Here elements of ${{\cal G}_{-}^2}^{\ast} $
of the form $L_{+} = \xi \, u_{-1} + u_0 + \xi^{-1} \, u_1 +
\xi^{-2} \, u_2 $,
span an invariant subspace under
$\d_{R_2} L_{+} = ad^{\ast}_{R_2} (X) L_{+}$ ,
i.e. they form a $R_2$-orbit of finite functional dimension $\, 4$ .
Defining the complete Lax operator $L^{(2)} = L_{+} + L_{-} = \xi \, u_{-1}
+ u_0 + \xi^{-1}\, u_1 + \xi^{-2} \, u_2 + \sum_{i \geq 3}
\xi^{-i} \, w_{i-3}\,$ we find from eq. \rf{rpb} the corresponding
Hamiltonian structure to be \ct{BAK85}:
$$
\left(\begin{array}{cc}
P^{(2)} & 0\\
0 & \; \O^{(2)} \end{array}
\right)
\;\; {\rm with } \;\;
P^{(2)}= \left(\begin{array}{cccc}
0 & 0 & 0 & - u_{-1} D + u_{-1}^{\pr} \\
0 & 0 & - u_{-1}D &  u_{-1} D^2 + u_0^{\pr} \\
0 & - D u_{-1} & 0 & D u_1\\
- D u_{-1} - u_{-1}^{\pr} & - D^2 u_{-1} - u_0^{\pr} & u_1 D & u_2 D +
D u_2 \end{array}
\right)
$$
where $P^{(2)}$ and $\O^{(2)}$ \rf{watform} are associated to
$\{ u_{-1}, u_0, u_1, u_2 \}$ and $\{ w_i \v i \geq 0\}$, respectively.
The LP structure $P^{(2)}$ is easily recognized as a semidirect product
of (centerless) Virasoro algebra generated by the spin $2$ field
$u_2\,$ with a subalgebra generated by the conformal fields
$\{ u_{-1}, u_0 + \pa_x u_{-1} , u_1 \}$ with spins $ -1,0,1$ ,
respectively.
Also, the LP structure with $\O^{(2)}$ \rf{watform}
corresponds to the centerless algebra ${\bf W_\infty^{\geq {\rm 3}}}$ which is
a subalgebra of $\Winf\,$ containing all generators
of spin $\geq 3\,$ .
\lskip
{\large{\bf 3. ``Gauge" Equivalence of Modified \KP Hierarchies to
Ordinary \KP}}
\lskip
{\bf 3.1 Ordinary Coadjoint Action on ${\bf R}$-Coadjoint Orbits as
Generalized Miura Transformation}

In this section we shall explicitly construct symplectic (hamiltonian)
maps among the various $R$-coadjoint orbits $\OR{}\,$ :
$\, \P : \OR{} \longrightarrow \ORti \;$ ,
where $\, R, {\wti R} = R_0 , R_1 ,
R_2 \,$ . The term ``symplectic" (``hamiltonian") means that under the
map $\P\,$, the LP bracket structure on $\ORti\,$
is transformed into the LP bracket structure on $\OR{}$ :
\be
\left\{ {\wti F}_1 , {\wti F}_2 \right\}_{\wti R}
\Bigl( \P (L)\Bigr)
= \left\{ {\wti F}_1 \Bigl( \P (L)\Bigr)\, , {\wti F}_2 \Bigl( \P (L)\Bigr)
\right\}_{R}  \lab{3-2}
\ee
where ${\wti F}_{1,2}\,$ are arbitrary functions on $\ORti$ and
we used notations $L\,$ and ${\wti L} = \P (L)\,$ for the
coordinates on $\OR{}$ and $\ORti$ , respectively.
As a consequence of \rf{3-2}, the infinite set of involutive integrals
of motion $\{ {\wti H}_N \lb {\wti L}\rb \}\,$ of the integrable system on
$\ORti\,$ are transformed into those of the integrable system on $\OR{}$
: $H_N \lb L\rb = {\wti H}_N \lb \P (L)\rb\,$ .

To this end we observe that the ordinary coadjoint actions of the Lie
algebra $\lie = \PsDA$ and the group $G= \PsDO\,$
on the dual space $\dlie = \dPsDA\,$
do {\em not} commute with any of the $R$-coadjoint actions \rf{R-coadj},
i.e. $ad^{\ast}(\cdot )$ and $Ad^{\ast}(\cdot )$ intertwine the orbits
for different $R$-coadjoint actions. Thus, it is natural to look for the
map $\, \P : \OR{} \longrightarrow \ORti \;$ in the form :
\be
{\wti L} \equiv \P (L) = Ad^{\ast} \Bigl( g(L) \Bigr) L  \lab{3-3}
\ee
where the group element $ g(L) \in \PsDO\,$ depends in general on the point
$L$ in $\OR{} \subset \dPsDA\,$ .

It is in the sense of eqs.\rf{3-2}-\rf{3-3} that
the integrable systems on the orbits $\OR{}$ for different $R$-matrices
are called ``gauge" equivalent. Also, from the point of view of
eq.\rf{3-2}, i.e. mapping of one Poisson bracket structure of an integrable
model into another one, the specific form of the $\PsDO$ group coadjoint
action \rf{3-3}, mapping $\OR{}$ into $\ORti$ , may be called
generalized Miura transformation.

Let us note the following important property of $Ad^{\ast} \Bigl( g(L)
\Bigr)\,$ in \rf{3-3}. It does not preserve the dual-projections
$P_{-}^{\ast}\, ,\, {\wti P}_{-}^{\ast}\,$ on $\OR{}$ and $\ORti$ ,
respectively : $\, {\wti P}_{+}^{\ast} Ad^{\ast} \Bigl( g(L) \Bigr)
P_{-}^{\ast} \neq 0 \,$ .

It is sufficient to prove ``gauge" equivalence for linear
functions on $\ORti$, namely:
\be
\lcurl \bil{{\wti L}}{{\wti X}} \, ,\, \bil{{\wti L}}{{\wti Y}}
\rcurl_{\wti R} \Bgv_{ {\wti L} = Ad^{\ast}\bigl( g(L)\bigr) L }
= \lcurl \bil{Ad^{\ast}\bigl( g(L)\bigr) L}{{\wti X}} \, ,\,
\bil{Ad^{\ast}\bigl( g(L)\bigr) L}{{\wti Y}} \rcurl_R  \lab{3-11}
\ee
Applying the general formula \rf{R-lie} to the r.h.s. of \rf{3-11} we
have :
\be
\lcurl \bil{\P (L)}{\wti X}\, ,\,\bil{\P (L)}{\wti Y} \rcurl_R =
- \llangle L \bgv \left\lb \nabla_L \bil{\P (L)}{\wti X}\, ,\,
\nabla_L \bil{\P (L)}{\wti Y} \right\rb_R \rrangle  \lab{3-12}
\ee
with
$$
\nabla_L \bil{\P (L)}{\wti X} = \( g^{-1}(L) \, {\wti X} \, g(L)\)
\xx
- \int dy {\rm Res}_\zeta \( \funcder{g(L)\yy}{L\xx} \,
g^{-1}(L)\yy \, \Bigl\lb {\wti X}\, ,\, \P (L) \Bigr\rb \yy \)
$$
\lskip
{\bf 3.2 ``Gauge" Transformation of \mKPa to Ordinary \KP}
\lskip
Let us first specialize eq.\rf{3-3} to the case
$\, \P : \OR{}\equiv {\cal O}(KP_{\ell =1}) \longrightarrow \ORti
\equiv {\cal O}(KP_{\ell =0})\;$ , i.e. :
\be
{\wti L} \equiv \xi + \suma{k=1} \xi^{-k} \, {\wti u}_k (x) =
\dAdB{0} \left( \xi + u_0 (x) + \xi^{-1} \, u_1 (x) +
\suma{k=2} \xi^{-k} \, v_{k-2} (x) \right)   \lab{3-4}
\ee
One easily finds the ``gauge" subgroup to be the Abelian group of
multiplication operators :
\be
g_0 (L) = \exp \p_0 (x) \;\;\;\;  ,\;\;\;\;   \pa_x \p_0 (x) = u_0 (x)
\lab{3-5}
\ee
by using the simple formula :
\be
e^{\p_0 (x)} \xi e^{- \p_0 (x)} = \xi - \pa_x \p_0 (x)   \lab{3-9}
\ee
Furthermore, from the structure of ${\wti L}\,$ \rf{3-4} we find that
only the $(+)\,$ parts of ${\wti X}$ and ${\wti Y}$ contribute in
\rf{3-12}, i.e. ${\wti X} \equiv {\wti X}_{\geq 0} \in
\lie^{\ell =0}_{+}$ \rf{subalg}. Finally, for $g(L) = g_0 (L)\,$
\rf{3-5}, we have :
\be
\funcder{g_0 (L)\yy}{L\xx} \, g_0^{-1}(L)\yy = -\h \vareps (x-y)
\xi^{-1}      \lab{var-L0}
\ee
Now specializing eq.\rf{3-12} by taking into account
\rf{3-4}-\rf{var-L0} and the form of the $R$-commutator \rf{R-bra} for
$\, \ell = 1$ , we obtain for ${\wti L} \equiv \exp \( \p_0 \) \, L
\, \exp \(-\p_0\) \,$:
\br
\lefteqn{
\lcurl \bil{{\wti L}}{\wti X}\, ,\,\bil{{\wti L}}{\wti Y}
\rcurl_{KP_{\ell =1}} = } \lab{3-14} \\
& -& \llangle {\wti L}\bgv
e^{\p_0} \, \biggl( \left\lb
\( e^{-\p_0}\, {\wti X}\, e^{\p_0} \)_{\geq 1} ,
\( e^{-\p_0}\, {\wti Y}\, e^{\p_0} \)_{\geq 1} \right\rb
- \left\lb
A_1 \( \p_0 , {\wti L} \) \( {\wti X} \) \, , \,
A_1 \( \p_0 , {\wti L} \) \( {\wti Y} \)
\right\rb   \biggr) \, e^{-\p_0} \rrangle  \nonu
\er
where $A_1 \( \p_0 , {\wti L} \) \( {\wti X} \) \equiv
( e^{-\p_0}\, {\wti X}\, e^{\p_0} )_{(0)} +
( \pa_x^{-1} \Res \lb {\wti X} , {\wti L} \rb ) \xi^{-1}$ and
the subscript $\scriptstyle{ (0)}$ means taking the zero-order part of the
$\xi$-expansion of the corresponding symbol. Note that in any term
on the r.h.s. of \rf{3-14} of the form $\,\left\langle {\wti L} \v {\wti Z}
\rrangle\,$ only the projection $\, {\wti Z}_{\geq -2}\,$ contributes.
Using the simple identity :
\br
e^{\p_0} \, \left\lb
\( e^{-\p_0}\, {\wti X}\, e^{\p_0} \)_{\geq 1} ,
\( e^{-\p_0}\, {\wti Y}\, e^{\p_0} \)_{\geq 1} \right\rb
 \, e^{-\p_0} & = &\left\lb {\wti X},{\wti Y} \right\rb \lab{3-15}\\
 &-& \left\lb \( e^{-\p_0}\, {\wti X}\, e^{\p_0} \)_{(0)} ,\,
{\wti Y} \right\rb - \left\lb {\wti X}\, ,
\( e^{-\p_0}\, {\wti Y}\, e^{\p_0} \)_{(0)} \right\rb
\nonu
\er
to rewrite the first commutator on the r.h.s. of \rf{3-14},
we easily find that the contribution of the terms in the second
commutator on the r.h.s. of \rf{3-14} are precisely canceled by the
second and the third term on the r.h.s. of \rf{3-15}. Thus, \rf{3-14}
reduces to the form :
\be
\lcurl \bil{{\wti L}(L)}{\wti X}\, ,\,\bil{{\wti L}(L)}{\wti Y}
\rcurl_{KP_{\ell =1}} =
- \llangle {\wti L}\bgv \left\lb {\wti X}, {\wti Y} \right\rb
\rrangle =
\lcurl \bil{{\wti L}}{\wti X}\, ,\,\bil{{\wti L}}{\wti Y}
\rcurl_{KP}   \lab{3-16}
\ee
which establishes the ``gauge" equivalence of \mKPa and \KP , i.e.
that the generalized Miura-like transformation \rf{3-4}-\rf{3-5}
maps the Poisson bracket structure of \KP into that of \mKPa and
vice versa.
\lskip
{\bf 3.3 ``Gauge" Transformation of \mKPb to Ordinary \KP}
\lskip
It is simpler to first establish the ``gauge" equivalence between
the modified $KP$ hierarchies \mKPb and \mKPa. The
desired result follows by combining the results of this and the
previous subsections.

Specializing eq.\rf{3-3} to the case
$\, \P : \OR{}\equiv {\cal O}(KP_{\ell =2}) \longrightarrow \ORti
\equiv {\cal O}(KP_{\ell = 1})\,$, we have :
\br
{\wti L} &\equiv &
\xi + {\wti u}_0 (x) + \xi^{-1} \, {\wti u}_1 (x) +
\suma{k=2} \xi^{-k} \, v_{k-2} (x)   \lab{3-6}\\
& =& \dAdB{1} \left( \xi \, u_{-1}(x) + u_0 (x) + \xi^{-1} \, u_1 (x) +
\xi^{-2} \, u_2 (x) + \suma{k=3} \xi^{-k} \, w_{k-3} (x) \right)
\nonu
\er
Here we find the ``gauge" subgroup to be the (centerless) Virasoro group :
\br
g_1 (L) = \exp \bigl( \p_1 (x) \xi \bigr) \phantb    \lab{3-7}  \\
u_{-1}\( F_{\p_1}(x)\) = \pa_x F_{\p_1}(x) \;\;\;\; {\rm with} \;\;\;\;
F_{\p_1}(x) \equiv \left( e^{\p_1 (x) \pa_x} x \right)  \lab{3-8}
\er
In \rf{3-8} $F_{\p_1}(x)\,$ denotes the global group Virasoro
diffeomorphism generated by the Virasoro algebra element $\p_1 (x) \xi
\simeq \p_1 (x) \pa_x\,$ . Note that all exponents involving symbols are
operator ones. To obtain \rf{3-6} one uses the simple formulas :
\be
e^{\p_1 (x) \xi}\, \xi\, e^{- \p_1 (x) \xi} =
{1\o \pa_x {F_{\p_1}(x)}} \, \xi \;\;\; ,\;\;\;
e^{\p_1 (x) \xi} \, u(x) \, e^{- \p_1 (x) \xi} = u \( F_{\p_1}(x) \)
\lab{3-10}
\ee
In the present case the analog of \rf{var-L0} reads :
\be
\funcder{g_1 (L)\yy}{L\xx} \, g_1^{-1}(L)\yy =
-\h \vareps \( x-F_{\p_1}(y)\)\, {1\o {u^2_{-1}(x)}} \,\xi^{-2}\,\zeta
\lab{var-L1}
\ee
Specializing formula \rf{3-12} yields for $ {\wti L} \equiv e^{\p_1 \xi}
L e^{-\p_1 \xi}$ :
\br
\lefteqn{
\lcurl \bil{{\wti L}}{\wti X}\, ,\,\bil{{\wti L}}{\wti Y}
\rcurl_{KP_{\ell =2}} =  -\llangle {\wti L} \bgv
e^{\p_1 \xi} \, \biggl( \left\lb
\( e^{-\p_1 \xi}\, {\wti X}\, e^{\p_1 \xi} \)_{\geq 2} ,
\( e^{-\p_1 \xi}\, {\wti Y}\, e^{\p_1 \xi} \)_{\geq 2} \right\rb
 \right.} \nonu\\
&-& \left. \left\lb
\( A_2 \( \p_1 , {\wti L} \) \( {\wti X} \) \)_{\leq 1} \, , \,
\( A_2 \( \p_1 , {\wti L} \) \( {\wti Y} \) \)_{\leq 1}
\right\rb \biggr) e^{-\p_1 \xi} \rrangle  \lab{3-17}
\er
where $ A_2 \( \p_1 , {\wti L} \) \( {\wti X} \)  =
e^{-\p_1 \xi} \( {\wti X} +
\( \pa_x^{-1} \lb {\wti X} , {\wti L} \rb_{(-2)} \) \xi^{-2}
\) e^{\p_1 \xi}$.
Similarly as in \rf{3-14}, in any term of the form
$\,\left\langle {\wti L} \v {\wti Z}\rrangle\,$ on the r.h.s. of
\rf{3-17} only the projection $\, {\wti Z}_{\geq -2}\,$ contributes.
The subscript $(-2)$ means taking the coefficient in front of $\xi^{-2}$
in the corresponding symbol expansion.

Noting that
$\, Ad \( g^{-1}_1 (L)\)\,$ preserves the splitting $\, {\wti X} =
{\wti X}_{\geq 1} + {\wti X}_{\leq 0}\,$ corresponding to \mKPa
\rf{subalg}, one can rewrite \rf{3-17} as :
\br
\lcurl \bil{{\wti L}(L)}{\wti X}\, ,\,\bil{{\wti L}(L)}{\wti Y}
\rcurl_{KP_{\ell =2}} =   - \llangle {\wti L}(L) \bgv
\left\lb {\wti X}_{\geq 1}, {\wti Y}_{\geq 1} \right\rb
- \left\lb {\wti X}_{\leq 0}, {\wti Y}_{\leq 0} \right\rb \rrangle
\lab{3-18}    \\
+ \llangle {\wti L}(L) \bgv \biggl\lb {\wti X} + \( \pa_x^{-1}
\left\lb {\wti X} ,{\wti L} \right\rb_{(-2)} \) \xi^{-2}\, , \,
e^{\p_1 \xi} \lcurl
\( e^{-\p_1 \xi} {\wti Y} e^{\p_1 \xi}\)_{(1)} \xi \rcurl e^{-\p_1 \xi}
\biggr\rb  \rrangle   \lab{3-19} \\
- \llangle {\wti L}(L) \bgv\biggl\lb {\wti Y} + \( \pa_x^{-1}
\left\lb {\wti Y} ,{\wti L} \right\rb_{(-2)} \) \xi^{-2}\, , \,
e^{\p_1 \xi} \lcurl
\( e^{-\p_1 \xi} {\wti X} e^{\p_1 \xi}\)_{(1)} \xi \rcurl e^{-\p_1 \xi}
\biggr\rb \rrangle      \lab{3-20}
\er
Now, accounting for the structure of $\, {\wti L}(L)\,$ \rf{3-6}, one
can easily show that both terms \rf{3-19} and \rf{3-20} vanish
separately. Thus, we are left with \rf{3-18} only, i.e. :
\be
\lcurl \bil{{\wti L}(L)}{\wti X}\, ,\,\bil{{\wti L}(L)}{\wti Y}
\rcurl_{KP_{\ell =2}} =
\lcurl \bil{{\wti L}}{\wti X}\, ,\,\bil{{\wti L}}{\wti Y}
\rcurl_{KP_{\ell =1}}   \lab{3-21}
\ee
which establishes the ``gauge" equivalence of \mKPb and \mKPa $\,$ , and
due to \rf{3-16}, also the ``gauge" equivalence of \mKPb to ordinary \KP .
\lskip
{\large {\bf 4. Applications. A New 4-Boson Representation of
$\,\Win1$ }}
\lskip
{\sl Lenard Relations :}
As mentioned in section {\em 2.1}
above, Lenard relations shown below eq.\rf{uflow} played
an important role in establishing the bi-Hamiltonian structure for the
ordinary $KP$ hierarchy.
Here we comment on how the ``gauge" equivalence
between the various $KP$ hierarchies carries the Lenard relations over to
the modified $KP$ hierarchies.
First we note that $\,\Tr {\wti L}^n =\Tr L^n\,$, with $L\,$ being a Lax
operator in the modified $KP$ hierarchy (notations as in
\rf{3-4},\rf{3-6}), follows from the ``gauge "equivalence
and ensures that the Hamiltonians $H_n$ remain identical for all \KPl
(upon extending $H_n$ as functions from orbits $\OR{}$ to the whole
dual space $\dlie$ ).
For simplicity we now discuss the case of \mKPa with
the Lax operator $L$ related through ${\wti L} \equiv \exp \( \p_0 \) \, L
\, \exp \(-\p_0\)$ to the Lax $ {\wti L} $ of usual $KP$ hierarchy.
One can easily show that the ordinary Lenard relations
$\{ H_n \,, \, {\wti L} \}_2= \{ H_{n+1} \,, \, {\wti L} \}_1$  translate
now to the new Lenard relations :
\be
 \{ H_{n+1} \,, \, L \}_1 + \lb  \{ H_{n+1} \,, \, \p_0 \}_1 \, , \, L \rb
 = \{ H_n \,, \, L \}_2 + \lb  \{ H_{n} \,, \, \p_0 \}_2 \, , \, L \rb
 \lab{nlenard}
\ee
Especially for the two-boson $R_1$-orbit $\, L_{+} =\xi + u_0 + \xi^{-1} u_1 $
we get
the relations $\{ H_n , u_i \}_2= \{ H_{n+1} , u_i \}_1$  with $i=0,1$
reproducing the second bracket structure found in \ct{BAK85,2boson}.
\lskip
{\sl New Representations of W-algebras :}
Here we will use the symplectic ``gauge" equivalence map to construct new
representation of the $\Win1\,$ algebra.
Let us first recall eq. \rf{3-4} and solve for coefficients of the Lax
operator on the l.h.s. in terms of the coefficients given on the r.h.s.
of this equation. One easily finds :
\be
{\wti u}_{k+1} = u_1 P_k (u_0) + \sum_{n=2}^{k+1}
{ k \choose n-1} v_{n-2} P_{k+1-n} (u_0) \qquad \; k \geq 0
\lab{nfaa}
\ee
where $P_k (u_0 ) \equiv \( \pa + u_0 \)^k \cdot 1$, are
the so called Fa{\'a}
di Bruno polynomials and the fields on the r.h.s. satisfy the LP
bracket structure described in \rf{ham1}.
As a corollary of the symplectic character of the ``gauge"
transformation we conclude that ${\wti u}_{k+1}\,$ \rf{nfaa} satisfy the
Poisson-bracket $\Win1\,$ algebra described by the form $\O^{(0)}$
{}from \rf{watform} (note that the index labeling $\, {\wti u}_{k+1}\,$
is precisely equal to its conformal spin).
Specifically, putting in \rf{nfaa}
all $v_i$ to zero we recover the two-boson representation
${\wti u}_{k+1} = u_1 P_k (u_0)\,$
of $\Win1\,$ algebra described in \ct{BAK85,2boson}(see also \ct{wuyu} for
another related two-boson representation).
The semiclassical limit is simply obtained by taking $P_k (u_0) \to
u_0^k$ in \rf{nfaa} and yields the generators of ${\bf w_{1+\infty}}$
algebra.

Similar considerations applied to \mKPb$\;$ result in a
new non-standard 4-boson representation of $\Win1\,$ algebra.
Indeed, performing a ``gauge" transformation consisting of a
composition of \rf{3-4} and \rf{3-6} on the 4-boson $R_2$-orbit
$\, L_{+} = \xi \, u_{-1} + u_0 + \xi^{-1}\, u_1 + \xi^{-2}\, u_2 \,$
with $\, \p_1\,$ as in \rf{3-8} and $\, \pa_x \p_0 (x) = \Bigl( u_0 +
\pa u_{-1} \Bigr) (F_{\p_1}(x))\,$, we obtain the following \KP
$\, {\wti L}\,$ operator :
\br
{\wti L} = e^{\p_0}\, e^{\p_1 \xi} \Bigl(
\xi \, u_{-1} + u_0 + \xi^{-1}\, u_1 + \xi^{-2}\, u_2 \Bigr)
e^{-\p_1 \xi}\, e^{-\p_0}  =
\xi + \suma{k \geq 1} \xi^{-k}\, U_k    \lab{4-gauge}  \\
U_{k+1} \equiv
{\hat u}_1 P_k \( {\hat u}_0 \) +
{\hat u}_2 Q_{k-1} \( {\hat u}_0 \, ,\, {\hat v}_0 \)
\qquad\qquad \lab{4boson}
\er
using notations :
\br
{\hat u}_0 (x) \equiv \Bigl( u_0 + \pa u_{-1} \Bigr) \( F_{\p_1}(x)\)
\;\;\;\; , \;\;\;\;
{\hat v}_0 (x) \equiv \pa_x \ln \Bigl( \pa_x F_{\p_1} \Bigr) \lab{u0-v0}   \\
{\hat u}_1 (x) \equiv \pa_x F_{\p_1} u_1 \( F_{\p_1}(x)\) \;\;\;\; ,\;\;\;\;
{\hat u}_2 (x) \equiv \( \pa_x F_{\p_1} \)^2 u_2 \( F_{\p_1}(x)\)
\lab{u1-u2}  \\
Q_k \({\hat u}_0 \, ,\,{\hat v}_0 \) \equiv \sum_{s=0}^k
\(\pa_x + {\hat u}_0 + {\hat v}_0 \)^{k-s}
\(\pa_x + {\hat u}_0 \)^s \cdot 1 \;\;\;\;\;\; (\; k \geq 0\; ) \lab{biFaa}
\er
Again, because of the symplectic property of the ``gauge" transformation
\rf{4-gauge} the fields $\, U_k\,$ \rf{4boson} span a
$\,\Win1\,$ LP bracket algebra realized in terms of the four bosonic
fields \rf{u0-v0} and \rf{u1-u2}. The Poisson bracket algebra of the
latter :
\br
\lcurl {\hat u}_1 (x) , {\hat u}_0 (y) \rcurl = - \pa_x \d (x-y)
\lab{4pb-a}   \\
\lcurl {\hat u}_2 (x) , {\hat v}_0 (y) \rcurl = - {\hat v}_0 (x)\,
\pa_x \d (x-y) + \pa_x^2 \d (x-y)  \lab{4pb-b}  \\
\lcurl {\hat u}_2 (x) , {\hat u}_2 (y) \rcurl =
- 2 {\hat u}_2 (x) \pa_x \d (x-y) - \pa_x {\hat u}_2 \, \d (x-y)
\qquad\qquad \lab{4pb-c}
\er
the rest being zero, is a direct sum of the Heisenberg algebra of
$\, \({\hat u}_0 ,{\hat u}_1 \) \,$ with the conformal algebra of spin
$\, 2\,$ and non-primary spin $\, 1\,$ fields
$\, \({\hat u}_2 ,{\hat v}_0 \) \,$.
Let us point out that the deformation of the conformal algebra
\rf{4pb-b}-\rf{4pb-c} with $\, \lcurl {\hat v}_0 (x) , {\hat v}_0 (y)
\rcurl = -c \pa_x \d (x-y)\,$ already appeared in the
construction \ct{2boson} of the 2-boson representation of the second
bracket structure of \KP ,
while the fields $\, \({\hat u}_0 ,{\hat u}_1 \) \,$ comprise
the usual two-boson content of $\Win1\,$ representation
(the first term on the r.h.s. of \rf{nfaa}).
The above 4-boson construction brings these two Bose structures together
to yield a new representation \rf{4boson} of the $\Win1\,$ algebra.

It is an interesting problem to study the quantization of the integrable
system corresponding to the 4-boson realization of \KP $\,$
(quantization of the 2-boson realization of \KP has been already
undertaken in \ct{west}).
\lskip
{\bf Acknowledgements.} $\;$ H.A., E.N and S.P. thank S. Solomon for
hospitality at the Hebrew University of Jerusalem.
Support for H.A. by {\sl U.S.-Israel BSF} is gratefully
acknowledged. I.V. is very indebted to A. Radul for illuminating
correspondence. S.P. and E.N. are thankful to CERN Theory Division
and J. Ellis for hospitality during the final stage of the present work.

\end{document}